\pdfoutput=1
\documentclass[aps,prx,twocolumn,amsmath,amssymb,floatfix]{revtex4-1}

\usepackage{physics}
\usepackage{graphicx}  
\usepackage{dcolumn}   
\usepackage{bm}        
\usepackage{amssymb}   
\usepackage{amsmath}
\usepackage{mathrsfs}
\usepackage{epigraph}
\usepackage{braket}
\usepackage{gensymb}
\usepackage{lmodern}
\usepackage{lipsum}
\usepackage{marvosym}
\usepackage{ tipa }
\usepackage{bbold}
\usepackage{dsfont}
\usepackage{esint}
\usepackage{mathdots}
\usepackage{xcolor}
\usepackage{appendix}
\usepackage[colorlinks = true,
linkcolor = blue,
urlcolor  = blue,
citecolor = blue,
anchorcolor = blue]{hyperref}
\usepackage{natbib}
\usepackage{multirow}
\usepackage[mathscr]{euscript}

\begin{document}
	
	\title{Spin pumping from antiferromagnetic insulator spin-orbit-proximitized by adjacent heavy metal: A first-principles Floquet-nonequilibrium Green's  function study} 
	
	\author{Kapildeb Dolui}
	\affiliation{Department of Physics and Astronomy, University of Delaware, Newark, DE 19716, USA}
	\author{Abhin Suresh}
	\affiliation{Department of Physics and Astronomy, University of Delaware, Newark, DE 19716, USA}
	\author{Branislav K. Nikoli\'{c}}
	\email{bnikolic@udel.edu}
	\affiliation{Department of Physics and Astronomy, University of Delaware, Newark, DE 19716, USA}
	
	\begin{abstract}
		Motivated by recent experiments [P. Vaidya {\em et al.}, Science {\bf 368}, 160 (2020)] on spin pumping from sub-THz radiation-driven uniaxial antiferromagnetic insulator (AFI) MnF$_2$ into  heavy metal (HM) Pt hosting strong spin-orbit (SO) coupling, we compute and compare pumped spin currents 
		in Cu/MnF$_2$/Cu and Pt/MnF$_2$/Cu heterostructures. Recent theories of spin pumping by AFI have relied on simplistic Hamiltonians (such as tight-binding) and the  scattering approach to quantum transport yielding the so-called interfacial spin mixing conductance (SMC), but the concept of SMC ceases to be applicable when SO coupling is present directly at the interface. In contrast, we use more general first-principles quantum transport approach which  combines noncollinear density functional theory  with Floquet-nonequilibrium Green's functions in order to take into account: {\em SO-proximitized AFI} as a new type of quantum material, different from isolated AFI and brought about by AFI hybridization with adjacent HM layer; SO coupling at interfaces; and evanescent wavefunctions penetrating from Pt or Cu into AFI layer to make its interfacial region {\em conducting} rather than insulating as in the original AFI. The DC component of pumped spin current $I_\mathrm{DC}^{S_z}$ vs. precession cone angle $\theta_{\vb*{l}}$ of the N\'{e}el vector $\vb*{l}$ of AFI {\em does not} follow putative  $I^{S_z}_\mathrm{DC} \propto \sin^2 \theta_{\vb*{l}}$, except for very small angles $\theta_{\vb*{l}} \lesssim 10^\circ$ for which we can define an {\em effective} SMC from the prefactor and find that it doubles from MnF$_2$/Cu to MnF$_2$/Pt interface. In addition, the angular dependence $I^{S_z}_\mathrm{DC}(\theta_{\vb*{l}})$ differs for opposite directions of precession of the N\'{e}el vector, leading to twice as large SMC for the right-handed than for the left-handed chirality of the precession mode. 
	\end{abstract}
	\maketitle
	%===================================================================================
	%\section{Introduction}\label{sec:intro}
	%===================================================================================
	\section{Introduction}\label{sec:intro}
	
	The spin pumping is a phenomenon in which precessing magnetization of a layer of metallic or insulating ferromagnet (FM), driven by $\sim$ GHz microwaves under the ferromagnetic resonance conditions, emits pure spin current into adjacent normal metal (NM) layers which is proportional to the frequency $\omega$ of precession~\cite{Tserkovnyak2005,Tatara2019}. It is analogous to adiabatic quantum pumping of charge~\cite{Switkes1999,Brouwer1998} in nanostructures with time-dependent potential and no bias voltage applied which, however, requires quantum coherence at ultralow temperatures. In contrast, spin pumping by magnetization dynamics is a robust and ubiquitous phenomenon observed in numerous spintronic devices at room temperature. The reason for this is that spin pumping  stems from processes at the FM/NM interface~\cite{Chen2009} that is always thinner than the phase coherence length. It can be accompanied by  pumped charge current as well, on the proviso that the left-right symmetry of the device is {\em broken} statically~\cite{Chen2009,Bajpai2019} or if spin-orbit (SO) coupling is present at the FM/NM interface~\cite{Mahfouzi2012,Mahfouzi2014,Chen2015,Ahmadi2017} or in the bulk~\cite{Varella2021} of FM layer, with DC component of pumped charge current scaling with precession frequency as  $\propto \omega^2$~\cite{Chen2009,Bajpai2019,Vavilov2001,FoaTorres2005} in the former or as $\propto \omega$~\cite{Mahfouzi2012,Mahfouzi2014} in the latter case. 

	\begin{figure}[ht!]
		\centering
		\includegraphics[width=0.49\textwidth]{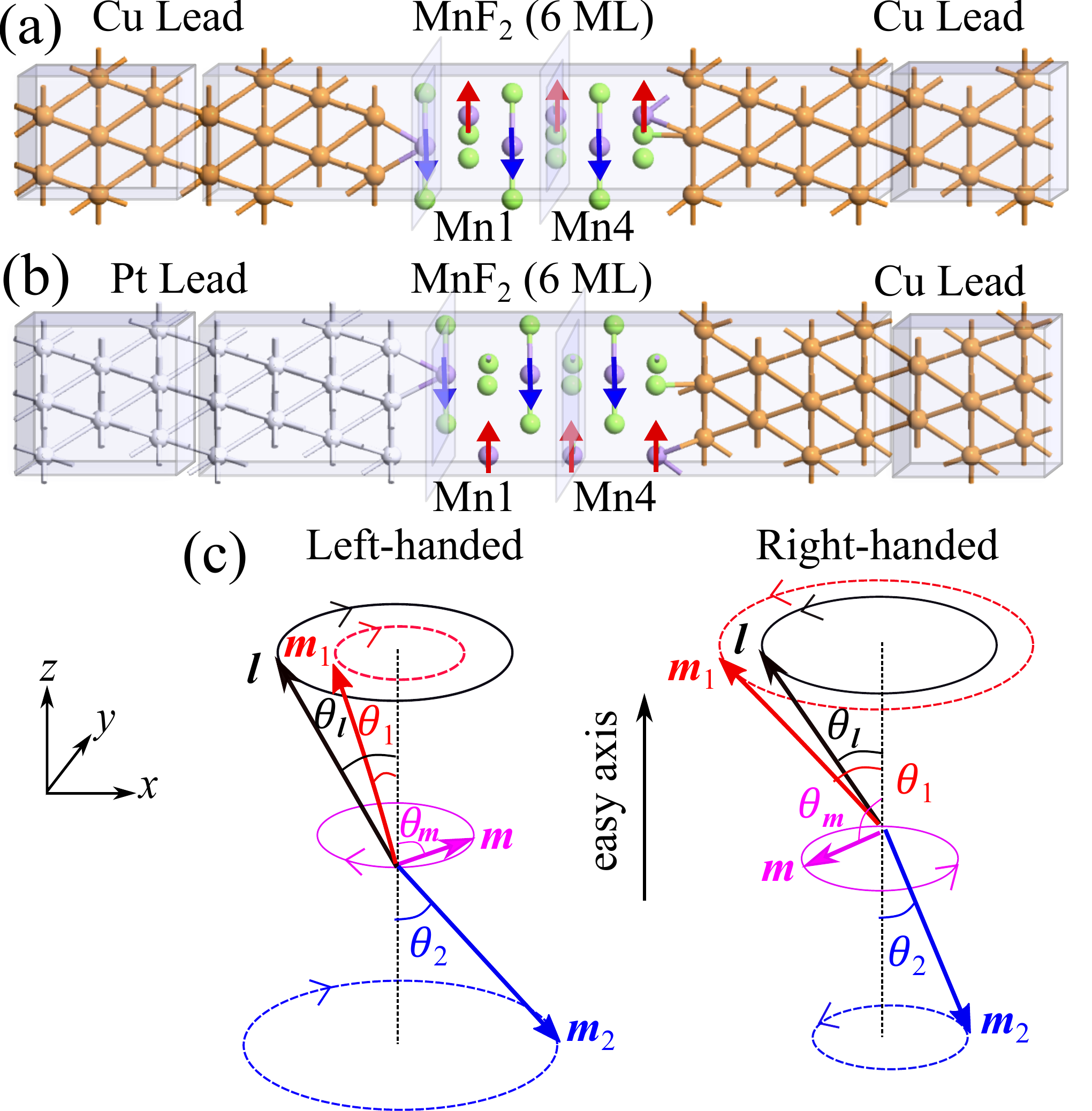}
		\caption{Schematic view of (a) Cu/AFI/Cu and (b) Pt/AFI/Cu heterostructures, where AFI layer is composed of six MLs of MnF$_2$ (001) with magenta spheres representing Mn atoms and light green spheres representing F atoms. The MnF$_2$ layer is sandwiched between semi-infinite Cu(111) and/or Pt(111) leads, while its  LMMs on the two sublattices are precessing with frequency $\omega$. The heterostructures are assumed to be infinite in the transverse direction, so that the depicted supercells are periodically repeated within the $yz$-plane. The semi-infinite leads terminate into macroscopic reservoirs without any bias voltage being applied between them. (c) Illustration of two precession modes~\cite{Vaidya2020,Cheng2014} around the easy-axis, with the left-handed or the right-handed chiralities, of sublattice magnetizations $\vb*{m}_1$ and $\vb*{m}_2$ at AF resonance  with two different cone angles $\theta_1$ and $\theta_2$. The same two modes  can also be described~\cite{Vaidya2020,Cheng2014}  by the corresponding precession of the N\'{e}el vector $\vb*{l}$ and net magnetization $\vb*{m}$ with cone angles  $\theta_{\vb*{l}}$ and $\theta_{\vb*{m}}$, respectively.}
		\label{fig:fig1}
	\end{figure}

    The recent interest in antiferromagnets (AFs) as possible active elements of spintronic devices~~\cite{Baltz2018,Jungwirth2016,Zelezny2018,Jungfleisch2018} has also led to intense efforts to demonstrate spin pumping from the dynamics of their localized magnetic moments (LMMs). By overcoming various obstacles---such as finding continuous source of THz radiation to force AF LMMs into precession; using antiferromagnetic insulators (AFIs) to exclude  large  rectification effect from AF metal itself; and using  AFIs with uniaxial anisotropy and the corresponding two well-defined precession modes around the easy axis---recent experiments have demonstrated spin pumping from  MnF$_2$~\cite{Vaidya2020} or Cr$_2$O$_3$~\cite{Li2020}. The pumped spin current from either of these two AFIs is directed into the adjacent heavy metal (HM) like Pt whose strong SO coupling  converts it into a charge signal by the inverse spin Hall effect (ISHE)~\cite{Saitoh2006,Johansen2017}. These advances open new avenues for spintronics  at ultrafast time scales and high frequencies which exceed by several orders of magnitude frequencies encountered in the dynamics of conventional FMs. 
    
    Let us recall that LMMs within FMs are aligned to generate nonzero magnetization in equilibrium below the Curie temperature. Their dynamics is dominated by crystalline anisotropy and applied magnetic fields, which leads to excitation frequencies of the order of $\sim$ GHz and it requires high energy for manipulating magnetization on time scales smaller than $\sim 1$ ns. Conversely, LMMs within collinear AFs point in alternating directions, so that net magnetization vanishes in equilibrium. Their dynamics is dominated by strong exchange interaction, required for relative canting of neighboring LMMs, which leads to excitation frequencies from $\sim 0.1$ THz (i.e., sub-THz)~\cite{Vaidya2020,Li2020} to $\sim 1$ THz~\cite{Jungfleisch2018}. Furthermore, changing chirality of precession of LMMs [Fig.~\ref{fig:fig1}(c)] leads to different directions~\cite{Cheng2014} of pumped spin current and the corresponding polarity of ISHE voltage~\cite{Saitoh2006,Johansen2017}, as confirmed in experiments~\cite{Vaidya2020} on MnF$_2$ by using  differently circularly polarized sub-THz radiation.
    
 	Collinear AFs can be viewed as being composed of two sublattices with their respective magnetizations pointing along the unit vectors $\vb*{m}_1$ and $\vb*{m}_2$. If one views na\"{i}vely spin pumping by precessing $\vb*{m}_1(t)$ and $\vb*{m}_2(t)$  as a phenomenon originating from two independent FMs,  one arrives at an incorrect conclusion that the net pumped spin current will be zero. However, careful theoretical analysis~\cite{Cheng2014}, based on the scattering approach to quantum transport~\cite{Tserkovnyak2005}, shows that spin pumping by sublattice magnetizations can add constructively to produce net nonzero outflowing spin current. In addition, spin pumping from one to another of the two sublattices of AFs can substantially modify the Landau-Lifshitz-Gilbert  dynamics of $\vb*{m}_1(t)$ and $\vb*{m}_2(t)$~\cite{Liu2017,Kamra2017,Yuan2020}. 
 	
 	The vector of pumped spin current outflowing into adjacent NM layer is given in the scattering approach~\cite{Tserkovnyak2005,Brouwer1998} by the following formula~\cite{Cheng2014}
 	\begin{equation}\label{eq:Is}
 		\mathbf{I}^{S}(t) = (I^{S_x}, I^{S_y}, I^{S_z}) = \frac{\hbar}{4\pi} [g_r( \vb*{l} \times \dot{\vb*{l}} + \vb*{m} \times \dot{\vb*{m}}) - g_i \dot{\vb*{m}}].
 	\end{equation}
    Here the N\'{e}el vector $\vb*{l} = (\vb*{m}_1 - \vb*{m}_2)/2$  specifies the direction of the staggered magnetization;  
 	$\vb*{m}$ specifies the direction of the net magnetization $\vb*{m} = (\vb*{m}_1 + \vb*{m}_2)/2$, where $\vb*{m} \equiv 0$ in equilibrium; and we use the shorthand notation $\dot{\vb*{l}} \equiv d\vb*{l}/dt$ and $\dot{\vb*{m}} \equiv d\vb*{m}/dt$. Equation~\eqref{eq:Is} introduces the real  $g_r$ and the imaginary $g_i$ part  ($g=g_r+ig_i$) of the so-called spin mixing conductance (SMC)~\cite{Tserkovnyak2005}. By taking the time-average of $\mathbf{I}^{S}(t)$ over one period of precession, only the first two terms in Eq.~\eqref{eq:Is} survive to contribute to the DC component of spin current $I^{S_z}_\mathrm{DC}$. Although $|\vb*{m}(t)| \ll |\vb*{l}(t)|$, the contribution of $\vb*{m} \times \dot{\vb*{m}}$ term to $I^{S_z}_{\mathrm{DC}}$ can be comparable to that of $\vb*{l} \times \dot{\vb*{l}}$ term because the two terms scale as $\propto \sin^2 \theta$ with their respective precession cone angles,  $\theta_{\vb*{m}}$ and $\theta_{\vb*{l}}$, where $\theta_{\vb*{l}} \rightarrow 0$ is small (for experimentally employed radiation power) while  $\theta_{\vb*{m}} \rightarrow 90^\circ$  is much larger  [Fig.~\ref{fig:fig1}(c)].

 	The SMC 
 	\begin{equation}\label{eq:smc}
 		g = \sum_{nm} [\delta _{n m} - (r_{n m}^{\uparrow\uparrow})(r_{n m}^{\downarrow\downarrow})^{*}], 
 	\end{equation}
 	describes the transport of electronic spin as it reflects from the NM/FM interface~\cite{Tserkovnyak2005} where  $r^{\sigma\sigma}_{n m}$ is the reflection probability amplitude for an electron in incoming conducting channel  $m$ with spin $\sigma = \uparrow , \downarrow$ to end up in outgoing channel $n$ with spin $\sigma$. For example, in Ref.~\cite{Cheng2014} SMC was computed for the NM/AFI interface by considering semi-infinite NM layer and only one monolayer (ML) of AFI attached to it, as described by simplistic tight-binding (TB) Hamiltonian, thereby assuming that electrons incident from NM layer {\em do not} penetrate beyond one ML of AFI because of its insulating nature.
 	
     However, longer penetration depth is likely in realistic materials~\cite{Zutic2019}, meaning that several MLs of AFI can become conducting due 
     to doping by evanescent wavefunctions originating from the NM layer. This conjecture is proved below by spectral functions of MnF$_2$ ML in direct contact with Cu [Fig.~\ref{fig:fig2}(a)] or  with Pt [Fig.~\ref{fig:fig2}(e)], where states {\em emerge} in the gap of AFI. These states diminish but remain nonzero  at the Fermi level $E-E_F=0$ {\em even} within the 4th ML [Figs.~\ref{fig:fig2}(b) and \ref{fig:fig2}(f)] of MnF$_2$ away from the interface. Thus, all such MLs will couple to conduction electrons, so that dynamics of their LMMs can~\cite{Chen2009} contribute to pumped spin current. Most importantly, when NM harbors strong SO coupling, as is the case of HM Pt or Ta layers employed  experimentally~\cite{Vaidya2020,Li2020} in order to convert pumped spin current into charge signal via ISHE~\cite{Saitoh2006,Johansen2017}, then SMC scattering formula becomes {\em inapplicable}~\cite{Tserkovnyak2005,Mahfouzi2012,Mahfouzi2014,Dolui2020,Chen2015,Ahmadi2017,Liu2014a} because interfacial SO coupling will generate nonzero elements, $r_{n m}^{\uparrow\downarrow} \neq 0$ and $r_{n m}^{\downarrow\uparrow} \neq 0$, of the scattering matrix. 
     
     In addition, few MLs of AFI can become {\em SO-proximitized}~\cite{Zutic2019}, akin to metallic FMs  proximitized by HMs~\cite{Dolui2017}, topological insulators~\cite{Marmolejo-Tejada2017}, and two-dimensional materials~\cite{Dolui2020b}. This is confirmed by complex equilibrium spin textures [Figs.~\ref{fig:fig2}(c), ~\ref{fig:fig2}(d), ~\ref{fig:fig2}(g) and ~\ref{fig:fig2}(h)] within MnF$_2$ MLs embedded into Cu/MnF$_2$/Cu [Figs.~\ref{fig:fig1}(a)] or Pt/MnF$_2$/Cu [Fig.~\ref{fig:fig1}(b)]  heterostructures. Finally, first ML of Pt in direct contact with MnF$_2$ will acquire nonzero LMMs \mbox{$\approx 0.013$ $\mu_B$} (vs. \mbox{$\approx 4.45$ $\mu_B$} on Mn atoms,  where $\mu_B$ is the Bohr magneton) as the signature of the magnetic proximity effect. 
        
     In this study, we bypass simplistic TB models~\cite{Cheng2014} of AFI/NM interfaces by employing noncollinear density functional theory (ncDFT)~\cite{Capelle2001, Eich2013a} which accurately captures in Fig.~\ref{fig:fig2} {\em proximity band structure}~\cite{Marmolejo-Tejada2017}  around the AFI/NM interface. The inapplicability of SMC scattering formula~\cite{Tserkovnyak2005,Cheng2014}, due to strong interfacial SO coupling, is bypassed by sending ncDFT Hamiltonian into charge conserving solution~\cite{Mahfouzi2012,Mahfouzi2014,Dolui2020} of Floquet-nonequilibrium Green's function (Floquet-NEGF) formulas~\cite{Arrachea2006,Shevtsov2013} which nonperturbatively~\cite{Mahfouzi2012} (i.e., for arbitrary cone angle or frequency) evaluates the DC component of pumped spin current. 
     
     Such first-principles quantum transport methodology, developed recently~\cite{Dolui2020} for atomistically defined quantum systems driven by time-periodic fields, is  applied to  Cu/MnF$_2$/Cu and Pt/MnF$_2$/Cu heterostructures [Fig.~\ref{fig:fig1}] where precessing sublattice magnetizations of MnF$_2$ AFI layer pump spin current into semi-infinite Cu or Pt leads in the absence of any bias voltage between the macroscopic reservoirs into which the NM leads terminate. We use six MLs of MnF$_2$ AFI which make it possible to capture all relevant proximity-induced modification of AFI due to charge transfer and evanescent wavefunctions originating from the adjacent NM or HM layers, as well as from SO coupling at the interface (in the case of Cu) or both from the interface and the bulk (in the case of Pt) which propagate to some distance into AFI layer. 
     
     The paper is organized as follows. Section~\ref{sec:floquethamiltonian} explains how to construct time-independent Floquet Hamiltonian~\cite{Mahfouzi2012,Arrachea2006,Shevtsov2013} using standard static ncDFT calculations~\cite{Capelle2001,Eich2013a} for heterostructures hosting both magnetism and SO coupling. How to pass such Hamiltonian through Floquet-NEGF formalism and obtain pumped spin or charge currents from first-principles is explained in Sec.~\ref{sec:floquetnegf}, with the corresponding results for spin pumping by MnF$_2$/Cu or MnF$_2$/Pt interfaces given in Sec.~\ref{sec:smc}. The observed features of spin pumping by AFI are further clarified using simplistic one-dimensional (1D) model of the AFI/NM interface in Sec.~\ref{sec:tdnegf}.  Section~\ref{sec:spectral} describes proximity band structure~\cite{Marmolejo-Tejada2017,Dolui2017,Dolui2020b} around  AFI/HM or AFI/NM interfaces.  We conclude in Sec.~\ref{sec:conclusions} while also pointing at possible extensions of first-principles Floquet-NEGF formalism.
     
     \section{Models and methods}\label{sec:methods}
     
     \subsection{First-principles time-dependent and Floquet Hamiltonians from static ncDFT calculations}\label{sec:floquethamiltonian}
     
     The equilibrium single-particle spin-dependent Kohn-Sham (KS) Hamiltonian of heterostructures in Fig.~\ref{fig:fig1} is given by 
     \begin{eqnarray}\label{eq:ncdfthamiltonian}
     \mathbf{H}_\mathrm{KS} & = &  -\frac{\hbar^2 \nabla^2}{2m} + \mathbf{V}_\mathrm{H}(\mathbf{r}) + \mathbf{V}_\mathrm{ext}(\mathbf{r}) + \mathbf{V}_\mathrm{XC}(\mathbf{r}) \nonumber \\ 
     \mbox{} && + \mathbf{V}_\mathrm{SO}(\mathbf{r}) - {\bm \sigma} \cdot \mathbf{B}_\mathrm{XC}, 
     \end{eqnarray}
     within ncDFT~\cite{Capelle2001,Eich2013a}. Here ${\rm \mathbf{V}}_\mathrm{H}(\mathbf{r})$, ${\mathbf{V}}_\mathrm{ext}(\mathbf{r})$, and ${\mathbf{V}}_\mathrm{XC}(\mathbf{r}) = \delta E_\mathrm{XC}[n(\mathbf{r}),\mathbf{m}(\mathbf{r})]/\delta n(\mathbf{r})$ are  the Hartree potential, external potential and the exchange-correlation (XC) potential, respectively; $\mathbf{V}_\mathrm{SO}$ is additional potential due to SO coupling; ${\bm \sigma} = (\hat{\sigma}_x, \hat{\sigma}_y,\hat{\sigma}_z)$ is the vector of the Pauli matrices;  and the XC magnetic field, \mbox{$\mathbf{B}_\mathrm{XC}(\mathbf{r}) = \delta E_\mathrm{XC}[n(\mathbf{r}),\mathbf{m}(\mathbf{r})]/\delta \mathbf{m}(\mathbf{r})$}, is functional derivative with respect to the vector of magnetization density $\mathbf{m}(\mathbf{r})$. The extension of DFT to the case of spin-polarized systems is formally derived in terms of $\mathbf{m}(\mathbf{r})$ and total electron density $n(\mathbf{r})$. In the collinear DFT, $\mathbf{m}(\mathbf{r})$ points in the same direction at all points in space, while in ncDFT $\mathbf{m}(\mathbf{r})$ can point in an arbitrary direction~\cite{Capelle2001,Eich2013a}. The matrix representation of the XC magnetic field can be extracted from $\mathbf{H}_\mathrm{KS}$ matrix using   
     \begin{equation}\label{eq:bxc}
     \mathbf{B}_\mathrm{XC}=(2 \mathrm{Re}[\mathscr{H}^{\uparrow \downarrow}], -2\mathrm{Im}[\mathscr{H}^{\uparrow\downarrow}]  , \mathscr{H}^{\uparrow\uparrow} -  \mathscr{H}^{\downarrow \downarrow}),
     \end{equation}
    where 
    \begin{equation}\label{eq:hscr}
    \mathscr{H} = \mathbf{H}_\mathrm{KS} - \mathbf{V}_\mathrm{SO}.
    \end{equation}
    
    The {\em time-dependent} Hamiltonian of heterostructures in Fig.~\ref{fig:fig1} is constructed as  
    \begin{equation}\label{eq:tdhamiltonian}
    \mathbf{H}(t) = \mathbf{H}_0(\theta) + \mathbf{V}(\theta)e^{i\omega t} + \mathbf{V}^\dagger(\theta) e^{-i\omega t}, 
    \end{equation}
    where $\mathbf{H}_0$ and $\mathbf{V}$ are obtained from purely static ncDFT calculations as 
    \begin{equation}\label{eq:h0}
    \mathbf{H}_0(\theta) = \mathbf{H}_\mathrm{KS}(\theta) - \mathbf{V}(\theta) - \mathbf{V}^\dagger(\theta),
    \end{equation} and 
    \begin{equation}
     \mathbf{V}(\theta) = \frac{1}{4}\mathbf{B}_\mathrm{XC}^x(\theta)\otimes [ \hat{\sigma}_x - i\hat{\sigma}_y ].
    \end{equation} 
    The {\em time-independent} Floquet Hamiltonian~\cite{Shirley1965,Sambe1973} is then given by 
     \begin{equation}\label{eq:floquet_ham}
     	\check{\mathbf{H}}_\mathrm{F} = \begin{pmatrix}
     		\ddots & \vdots & \vdots & \vdots & \iddots\\
     		\cdots & \mathbf{H}_0 & \mathbf{V} & 0 & \cdots \\
     		\cdots & \mathbf{V}^\dagger & \mathbf{H}_0 & \mathbf{V} & \cdots \\
     		\cdots & 0 & \mathbf{V}^\dagger& \mathbf{H}_0 & \cdots \\
     		\iddots & \vdots & \vdots & \vdots & \ddots
     	\end{pmatrix},
     \end{equation}
     All matrices labeled as $\check{\mathbf{O}}$ are representations of operators acting in the Floquet-Sambe~\cite{Sambe1973} space, $\mathcal{H} =  \mathcal{H}_T \otimes \mathcal{H}_e$, where $\mathcal{H}_e$ is the Hilbert space of electronic states spanned by $|\phi_a\rangle$ and $\mathcal{H}_T$ is the Hilbert space of periodic functions with period $T=2\pi/\omega$ spanned by orthonormal Fourier vectors $\langle t|n \rangle = \exp(i n \omega t)$ where $n$ is integer. In this notation, $\check{\sigma}_\alpha = \bold{1}_T \otimes \hat{\sigma}_\alpha$ is the Pauli matrix in $\mathcal{H}$; $\bold{1}_T$ is the identity matrix in $\mathcal{H}_T$; and we also use below $\bold{1}$ as the identity matrix in $\mathcal{H}_e$.
    
     The ncDFT calculations on {\em nonperiodic} systems in Fig.~\ref{fig:fig1}, which eventually yield Hamiltonians $\mathbf{H}(t)$ [Eq.~\eqref{eq:tdhamiltonian}] and $\check{\mathbf{H}}_\mathrm{F}$ [Eq.~\eqref{eq:floquet_ham}], are performed using {\tt QuantumATK}~\cite{Smidstrup2019} package. We first employ the interface builder in the {\tt QuantumATK}~\cite{Smidstrup2019} package to construct a unit cell for the multilayer heterostructures in Fig.~\ref{fig:fig1} and use the experimental lattice constants of their respective materials, while the lattice strain at each interface is kept below 1.5$\%$. In order to determine the interlayer distance we carry out DFT calculations with the generalized gradient approximation (GGA) in the parametrization of Perdew, Burke and Ernzerhof (PBE)~\cite{Perdew1996}, as implemented in the \texttt{QuantumATK} package~\cite{Smidstrup2019}. 
     
\begin{figure*}[!ht]
		\centering
		\includegraphics[width=1.0\textwidth]{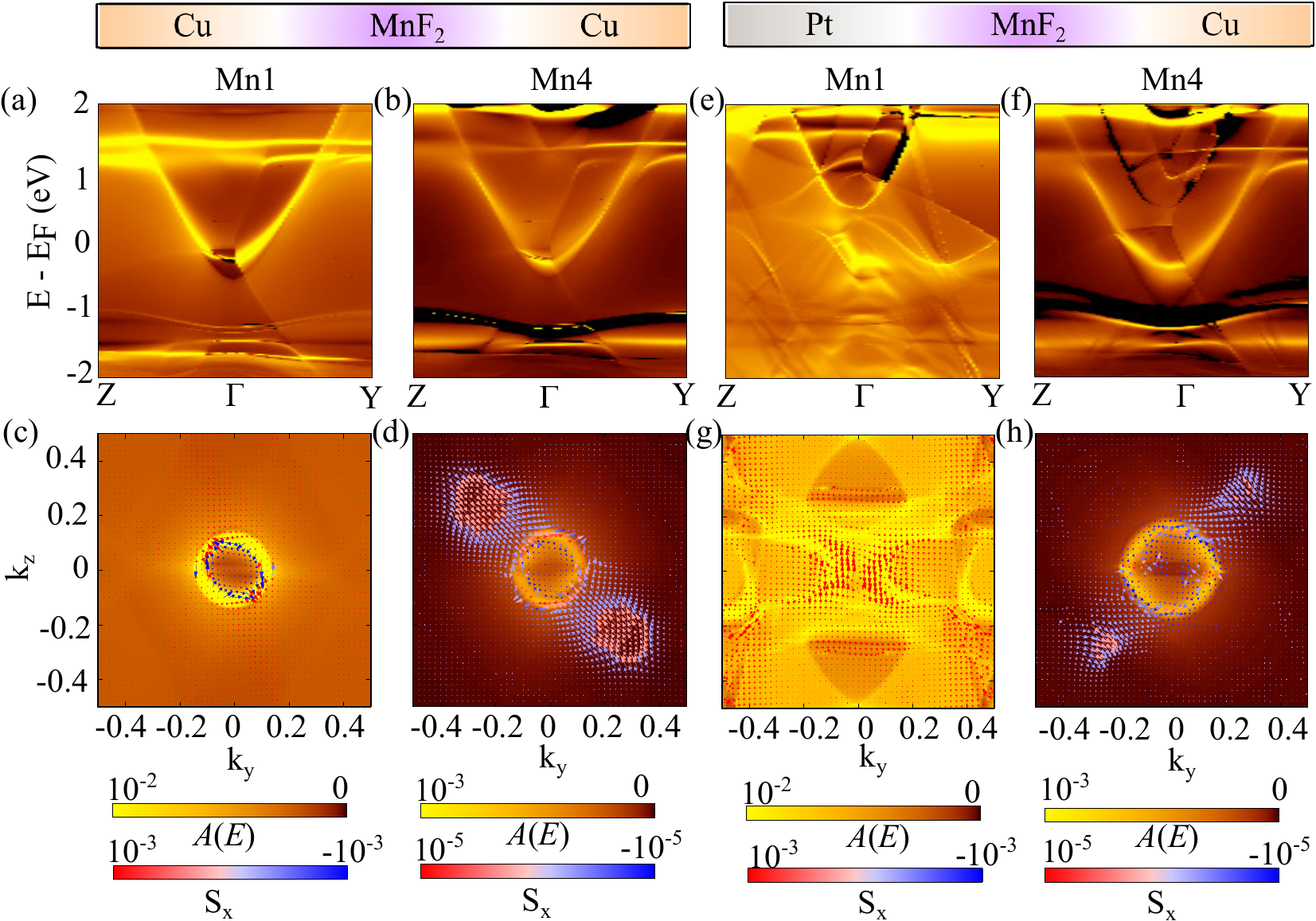}
		\caption{Spectral function $A(E;k_y,k_z, x\in\{{\rm Mn1,Mn4}\})$ plotted along the high-symmetry $k$-path, $Z-\Gamma-Y$ at the atomic plane passing through (a) Mn1 or (b) Mn4, as indicated in Fig.~\ref{fig:fig1}(a), for Cu/MnF$_2$/Cu  heterostructure with the N\'{e}el vector $\vb*{l}$ pointing along the $x$-axis. Panels (c)--(d) plot constant energy contours of the spectral function at the Fermi energy $E-E_F=0$, together with the corresponding spin textures where the out-of-plane spin expectation value $S_x$  is indicated in color (red for positive and blue for negative). Panels (e)--(h) show the same information as in respective panel (a)--(d), but for Pt/MnF$_2$/Cu heterostructure depicted in Fig.~\ref{fig:fig1}(b). The units for $k_y$ and $k_z$ are 2$\pi/b$ and 2$\pi/c$ where $b$ and $c$ are the lattice constants of the common unit-cell of the heterostructure.}
		\label{fig:fig2}
	\end{figure*}

     Prior to constructing Hamiltonians $\mathbf{H}(t)$ [Eq.~\eqref{eq:tdhamiltonian}] and $\check{\mathbf{H}}_\mathrm{F}$ [Eq.~\eqref{eq:floquet_ham}] 
     for the central region of Cu/MnF$_2$/Cu or Pt/MnF$_2$/Cu heterostructures---consisting of six MLs of MnF$_2$  with four MLs of Cu or Pt attached to each side, as illustrated in Fig.~\ref{fig:fig1}---we perform standard equilibrium Green's function (GF) calculations using \texttt{QuantumATK} package~\cite{Smidstrup2019} with PBE-GGA  exchange-correlation functional; norm-conserving pseudo-potentials for describing electron-core interactions; and ``SG15 (medium)" basis of localized orbitals~\cite{Schlipf2015}. Periodic boundary conditions are employed in the plane perpendicular to the transport direction ($x$-direction), with  9$\times$9 $k$-point grid for self-consistent calculation. The energy mesh cutoff for the real-space grid is 100 Hartree. These calculations make it possible to obtain the retarded GF in equilibrium
     \begin{equation}\label{eq:rgf} 
     	\mathbf{G}_0(E; \mathbf{k}_{\parallel}) = [E - \mathbf{H}_{\mathrm{KS}}(\mathbf{k}_{\parallel}) - \mathbf{\Sigma}_L^r(E,\mathbf{k}_{\parallel}) - \mathbf{\Sigma}_R^r(E,\mathbf{k}_{\parallel})]^{-1},
     \end{equation}
 	 where  $\mathbf{\Sigma}_{L,R}^r(E,\mathbf{k}_{\parallel})$ are the self-energies~\cite{Smidstrup2019,Rungger2008} due to the left ($L$) or the right ($R$) semi-infinite NM leads made of Cu or Pt. From $\mathbf{G}_0(E; \mathbf{k}_{\parallel})$ we extract the spectral function [Fig.~\ref{fig:fig2}] at an arbitrary plane  at position $x$ of nonperiodic systems in Fig.~\ref{fig:fig1} as
     \begin{equation}
     	A (E;\mathbf{k}_{\parallel};x) = -\frac{1}{\pi}\mathrm{Im}[\mathbf{G}_0(E;\mathbf{k}_{\parallel};x,x)].
     	\label{eq:spectral}
     \end{equation}
     Here the diagonal matrix elements $\mathbf{G}_{0}(E;\mathbf{k}_{\parallel};z,z)$ are computed by transforming the retarded GF from the local orbital to the  real-space representation, and \mbox{$\mathbf{k}_{\parallel} = (k_y, k_z)$} is the transverse $k$-vector. The constant energy contours of the spin-resolved spectral function at the chosen energy $E$ yield the spin textures [Figs.~\ref{fig:fig2}(c), ~\ref{fig:fig2}(d), ~\ref{fig:fig2}(g), ~\ref{fig:fig2}(h)] at position $x$.

     \subsection{Charge conserving solution of Floquet-NEGF equations for pumped currents}\label{sec:floquetnegf}
     
     The time-dependent NEGF formalism~\cite{Stefanucci2013,Gaury2014,Popescu2016} operates with two fundamental quantities~\cite{Stefanucci2013}---the retarded $\mathbf{G}^r(t,t')$ and the lesser $\mathbf{G}^<(t,t')$ GFs---which describe the density of available quantum states and how electrons occupy those states in nonequilibrium, respectively. They depend on two times, but solutions can be sought in other representations, such as the double-time-Fourier-transformed~\cite{Mahfouzi2012,Wang2003} GFs, ${\bf G}^{r}(E,E')$ and ${\bf G}^{<}(E,E')$. 
     
     In the case of periodic time-dependent Hamiltonians, they must take the form 
     \begin{equation}\label{eq:floquetgf}
     {\bf G}^{r,<}(E,E')={\bf G}^{r,<}(E,E+n\omega)={\bf G}^{r,<}_n(E),
     \end{equation} 
 	in accord with the Floquet theorem~\cite{Shirley1965,Sambe1973}. The coupling of energies $E$ and $E+ n\omega$ ($n$ is integer) indicates how `multiphoton' exchange processes contribute toward the pumped current. In the absence of many-body (electron-electron or electron-boson)  interactions, currents can be expressed using solely the Floquet-retarded-GF~\cite{Mahfouzi2012} 
 	\begin{equation}\label{eq:floquetrgf}
    [(E + \check{\bm \Omega})\check{\mathbf{S}} - \check{\mathbf{H}}_\mathrm{F} - \check{\mathbf{\Sigma}}^r(E) ]\check{\mathbf{G}}^r(E) = \check{\mathbf{1}},
    \end{equation} 
	which is composed of ${\bf G}^r_n(E)$ submatrices along the diagonal. Here 
	\begin{equation}\label{eq:checkomega}
		\check{\bm \Omega}=\mathrm{diag}[\ldots, -2\hbar\omega\mathbf{1}, -\hbar\omega\mathbf{1},0,\hbar\omega\mathbf{1},2\hbar\omega\mathbf{1},\ldots], 
	\end{equation}
	and 
	\begin{equation}\label{eq:checksigma}
		\check{\bm \Sigma}^r(E) = \mathrm{diag}[\ldots, {\bm \Sigma}^r(E-\hbar\omega),0, {\bm \Sigma}^r(E+\hbar\omega),\ldots],
	\end{equation}
	is the Floquet self-energy matrix. The submatrices \mbox{${\bm \Sigma}^r(E) = \sum_{p=L,R} {\bm \Sigma}_p^r(E)$} along the diagonal of $\check{\bm \Sigma}^r(E)$ are composed of standard~\cite{Rungger2008,Smidstrup2019} lead self-energies introduced in Eq.~\eqref{eq:rgf}. For nonorthogonal basis of localized orbitals $\ket{\phi_a}$,  $\check{\mathbf{S}}$ is an infinite matrix composed of overlap submatrices $\mathbf{S}$ along the diagonal with matrix elements $\mathrm{S}_{ab} = \braket{\phi_a|\phi_b}$. 

     In the adiabatic limit $\hbar\omega \ll E_\mathrm{F}$, justified by the Fermi energy $E_F \sim 1$ eV  of heterostructures in Fig.~\ref{fig:fig1} and  \mbox{$\omega \lesssim 1$ THz} precession frequency,  time-averaged (over a period $T$) pumped spin currents outflowing into the NM leads $p=L,R$ are given by~\cite{Mahfouzi2012}
     \begin{multline}\label{eq:floq_spj}
     	I_{p}^{S_\alpha} = \frac{\hbar}{4N_\mathrm{max}A_\mathrm{cell}} \int_\mathrm{BZ}   d\mathbf{k}_\parallel \, \mathrm{Tr}[\check{\sigma}_\alpha\check{\bold{\Gamma}}_p\check{\bold{\Omega}}\check{\bold{G}}^r\check{\bold{\Gamma}}\check{\bold{G}}^a \\ - \check{\sigma}_\alpha\check{\bold{\Gamma}}_p\check{\bold{G}}^r\check{\bold{\Gamma}}\check{\bold{\Omega}}\check{\bold{G}}^a ].
     \end{multline}
     Here \mbox{$\check{\bold{\Gamma}}_p(E) = i [\check{\bold{\Sigma}}_p^r(E) - (\check{\bold{\Sigma}}_p^r(E))^\dagger]$}; \mbox{$\check{\bold{\Gamma}}(E) = \sum_{p=L,R} \check{\bold{\Gamma}}_p(E)$}; and the Floquet-advanced-GF is \mbox{$\check{\bold{G}}^a(E) = [\check{\bold{G}}^r(E)]^\dagger$}. The pumped charge current is obtained from Eq.~\eqref{eq:floq_spj} by replacing $\check{\sigma}_\alpha \mapsto \bold{1}_T \otimes \hat{\sigma}_0$, where $\hat{\sigma}_0$ is the unit $2 \times 2$ matrix, and $\hbar/2 \mapsto e$.  Here all matrices depend on $\mathbf{k}_\parallel$ due to assumed periodicity of heterostructures depicted in Fig.~\ref{fig:fig1} within the $yz$-plane and absence of disorder, so that integration over the two-dimensional Brillouin zone (BZ) is performed and $A_\mathrm{cell}$ is the area of the unit cell in the transverse direction. In Fig.~\ref{fig:fig3}, we use \mbox{$k_y \times k_z =25\times25$} grid of $k$-points.

     Thus, Floquet-NEGF formalism replaces the original time-dependent NEGF~\cite{Varella2021,Gaury2014,Popescu2016} problem with  {\em time-independent} one at the cost of using infinite-dimensional matrices $\check{\mathbf{O}}$ due to infinite dimensionality of $\mathcal{H}_T$. However, in practice finite $|n| \le N_\mathrm{max}$ is chosen where this range can be expanded until the answer converges, thereby yielding a {\em nonperturbative} result. Note that trace in Eq.~\eqref{eq:floq_spj}, $\mathrm{Tr} \equiv  \mathrm{Tr}_e \mathrm{Tr}_T$, is summing over contributions from different subspaces of $\mathcal{H}_T$ so that the denominator includes $2N_\mathrm{max}$ to avoid double counting. The part of the trace operating in $\mathcal{H}_T$ space ensures that at each chosen $N_\mathrm{max}$ charge current is conserved, $I_\mathrm{L} \equiv I_\mathrm{R}$, unlike some other solutions~\cite{Wang2003,Kitagawa2011} of Floquet-NEGF equations where current conservation is ensured only in the limit $N_\mathrm{max} \rightarrow \infty$. In Fig.~\ref{fig:fig3} we use  $N_{\rm max}=2$,  determining the size of  truncated Floquet Hamiltonian matrix~\cite{Mahfouzi2012,Mahfouzi2014,Dolui2020}, which is sufficient to converge  within  1\%~\cite{Dolui2020}.

    \section{Results and discussion}
    
    \subsection{Proximity band structure around AFI/NM or AFI/HM interfaces}\label{sec:spectral}
	
	When AFI layer is attached to NM or HM layer, the evanescent wave functions from the latter can penetrate into the former while decaying exponentially away from the interface. For heterostructures in Fig.~\ref{fig:fig1} this picture is corroborated by comparing spectral function $A(E;\mathbf{k}_{\parallel})$  on the Mn1 monolayer [Figs.~\ref{fig:fig2}(a) and ~\ref{fig:fig2}(e)] directly at the interface with that on  
	Mn4 monolayer [Figs.~\ref{fig:fig2}(b) and ~\ref{fig:fig2}(f)], where both spectral functions exhibit nonzero states at the Fermi level $E-E_F=0$. This is in contrast to isolated AFI whose density of states is {\em zero} in the gap. 
	
	In addition,  evanescent wave functions can introduce proximity SO coupling~\cite{Marmolejo-Tejada2017,Dolui2017,Dolui2020b} into the AFI layer, as manifested by nonzero spin textures [Figs.~\ref{fig:fig2}(c), ~\ref{fig:fig2}(d),  ~\ref{fig:fig2}(g) and ~\ref{fig:fig2}(h)], with enhanced interfacial spin textures when Cu is replaced with Pt [compare Fig.~\ref{fig:fig2}(c) with Fig.~\ref{fig:fig2}(g)]. We note that nonzero spin textures in symmetric heterostructures NM/FM/NM or NM/AFI/NM, where NM layers and interfaces are identical on both sides, are forbidden~\cite{Dolui2017} due to preserved inversion symmetry, but left (Cu/MnF$_2$) and right (MnF$_2$/Cu) interfaces in the heterostructure in Fig.~\ref{fig:fig1}(a) are manifestly {\em not} identical.

	\begin{figure}
		\centering
		\includegraphics[width=0.49\textwidth]{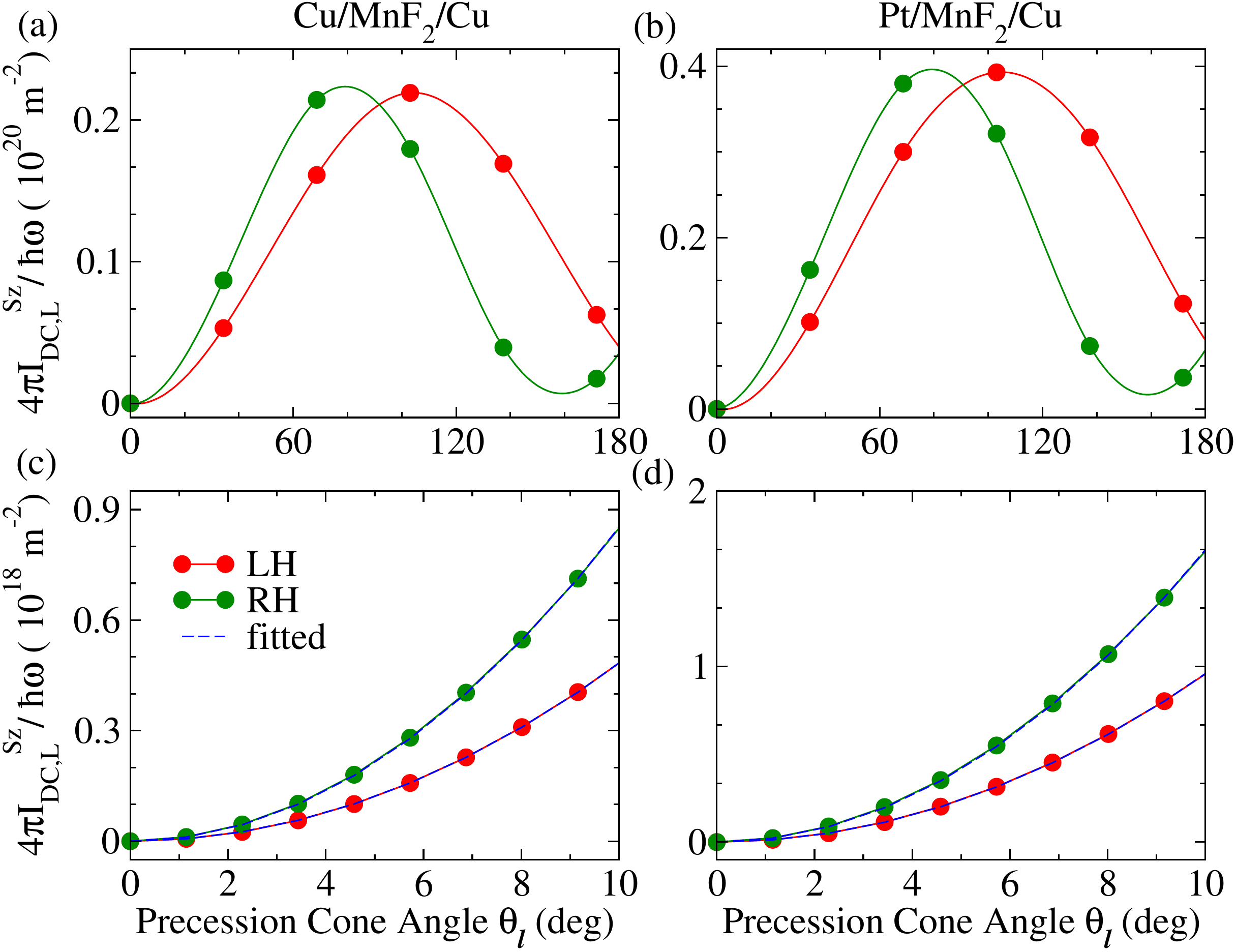}
		\caption{The angular dependence of the DC component of pumped spin current, $I_{\mathrm{DC},L}^{S_z}$ injected into the $L$ lead for LH (red circles + line) or  RH (green circles + line) AF precession mode [Fig.~\ref{fig:fig1}(c)] around the easy ($z$-axis) of MnF$_2$ AFI  within (a) Cu/MnF$_2$/Cu and (b) Pt/MnF$_2$/Cu heterostructures. The angle $\theta_{\vb*{l}}$ on the abscissa of all panels is precession cone angle of the N\'{e}el vector [Fig.~\ref{fig:fig1}(c)]. Panels (c) and (d) show the same information as in panels (a) and (b), respectively, but for small values of the precession cone angle $\theta_{\vb*{l}} \le 10^\circ$, where the curves are fitted with Eq.~\eqref{eq:effectivesmc} (blue dashed line) in order to extract {\em effective} SMC for AFI/NM and AFI/HM interfaces listed in Table~\ref{tab:smc}.  In all panels we use  $\theta_2/\theta_1 \approx (1+\sqrt{H_A/H_E})^2=1.29$~\cite{Cheng2014} for the LH mode of MnF$_2$ with $H_A/H_E = 0.018$~\cite{Vaidya2020} (or $\theta_1/\theta_2=1.29$ for the RH mode) and \mbox{$\hbar\omega = 10^{-3}$ eV}.}
		\label{fig:fig3}
	\end{figure}
	\begin{figure}[ht!]
		\centering
		\includegraphics[width=0.49\textwidth]{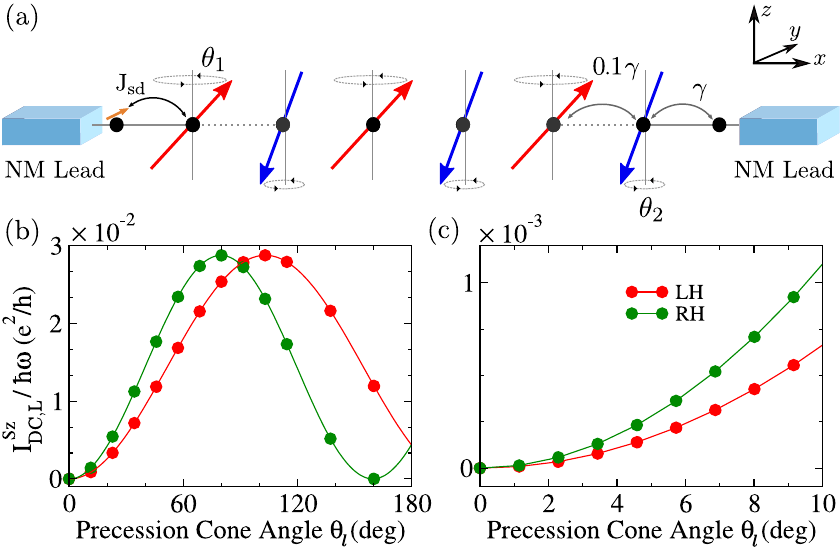}
		\caption{(a) Schematic view of a two-terminal device where 1D AFI is sandwiched between two semi-infinite NM leads (modeled as 1D TB chains) which terminate into macroscopic reservoirs kept at the same electrochemical potential. The AFI hosts six LMMs, denoted by red and blue arrows for two sublattices of AFI, which precess with frequency $\omega$ and precession cone angle $\theta_1$  and $\theta_2$, respectively with respect to the $z$-axis, with the ratio $\theta_1/\theta_2=1.29$ chosen the same as in Fig.~\ref{fig:fig3}. The conduction electrons hop from the NM leads into the AFI with hopping \mbox{$\gamma_{ij}=\gamma=1$ eV}, and they can penetrate up to the second site of AFI from the edge on each side where the hopping due to proximity effect of NM leads onto AFI is chosen as \mbox{$\gamma_{ij}=0.1 \gamma$}. The conduction electrons are brought out of equilibrium by precessing LMMs at two sites on each edge and  interact with LMMs via $sd$ exchange interaction of strength \mbox{$J_{sd} = 0.1 $ eV} [Eq.~\eqref{eq:tbh}]. (b) The angular dependence of  $I^{S_z}_{\mathrm{DC},L}$ with precession cone angle  $\theta_{\vb*{l}}$ [Fig.~\ref{fig:fig1}(c)] of the N\'{e}el vector for the LH (red circles + line) or the RH (green circles + line) mode. Panel (c) shows the same information as panel (b) but for small values of $\theta_{\vb*{l}} \le 10^\circ$.}
		\label{fig:fig4}
	\end{figure}

	\subsection{Spin pumping and effective spin mixing conductance from first-principles Floquet-NEGF formalism}\label{sec:smc}
	
	The presence of interfacial SO coupling can lead to spin memory loss~\cite{Dolui2017,Belashchenko2016,Kriti2020}, thereby substantially affecting proper extraction of spin transport-related parameters from spin pumping experiments~\cite{Rojas-Sanchez2014,Zhu2019}. But its effect on the magnitude of outflowing pumped spin current is not easy to conjecture since examples exist where the magnitude is enhanced~\cite{Chen2015,Jamali2015} or diminished~\cite{Dolui2020}. Therefore, we proceed to analyze angular dependence of the DC component of spin currents $I_{\mathrm{DC},L}^{S_z}$ outflowing into the $L$ lead in Fig.~\ref{fig:fig3}, where comparison can be made between the cases of weak SO coupling at Cu/MnF$_2$ [Fig.~\ref{fig:fig3}(a)] interface vs. strong SO coupling at Pt/MnF$_2$ [Fig.~\ref{fig:fig3}(b)] interface. This comparison, together with Table~\ref{tab:smc}, reveals  {\em enhancement} of pumped spin current with increasing interfacial SO coupling, by about a factor of two, upon switching from Cu to Pt layer. It is also closely related to the larger density of midgap states [Fig.~\ref{fig:fig2}(e)]  injected into MnF$_2$ by Pt. 
	
    Let us recall that at AF resonance two precession modes of sublattice magnetizations $\vb*{m}_1$ and $\vb*{m}_2$ are possible---with left-handed (LH) and right-handed (RH) chiralities~\cite{Cheng2014,Vaidya2020}---where both $\vb*{m}_1$ and $\vb*{m}_2$ undergo a clockwise or counterclockwise precession with $\pi$ phase difference, respectively, as illustrated in Fig.~\ref{fig:fig1}(c). The ratio of the precession cone angles $\theta_1$ and $\theta_2$ of  $\vb*{m}_1$ and $\vb*{m}_2$ depend on the ratio of anisotropy, $H_A$, and exchange, $H_E$, magnetic fields~\cite{Cheng2014}, which leads to the emergence of a small net magnetization $\vb*{m} = (\vb*{m}_1 + \vb*{m}_2)/2 \neq 0$ [Fig.~\ref{fig:fig1}(c)] in nonequilibrium. In Cu/MnF$_2$/Cu heterostructure with symmetric NM leads, the magnitude of $I_{\mathrm{DC},L}^{S_z}$ ($I_{\mathrm{DC},R}^{S_z}$) of  LH  mode is equal to that of $I_{\mathrm{DC},R}^{S_z}$~($I_{\mathrm{DC},L}^{S_z}$) of  RH mode. The dependence of $I_{\mathrm{DC},L}^{S_z}$ on the chirality of the precession mode in Fig.~\ref{fig:fig3} is in full accord with spin pumping experiments on  MnF$_2$/Pt bilayers~\cite{Vaidya2020}. 
    
    The Floquet-NEGF-computed pumped spin currents $I_{\mathrm{DC},L}^{S_z}$ in Figs.~\ref{fig:fig3}(a) and ~\ref{fig:fig3}(b) {\em do not} follow $\propto \sin \theta^2_{\vb*{l}}$ scaling with the precession cone angle $\theta_{\vb*{l}}$ of the N\'{e}el vector, in contrast to prediction~\cite{Chen2015} of Eq.~\eqref{eq:Is}. Nevertheless, since $I_{\mathrm{DC},L}^{S_z}  \propto \sin \theta^2_{\vb*{l}}$ at sufficiently small $\theta_{\vb*{l}} \lesssim 10^\circ$ (angles controlled by input radiation power can be tuned up to $\lesssim 20^\circ$~\cite{Jamali2015,Fan2010} without introducing nonlinearities), this makes it possible to define an {\em effective} SMC $g_{r}^{\mathrm{eff}}$ from the prefactor
    \begin{equation}\label{eq:effectivesmc}
    	I^{S_z}_{\mathrm{DC},L}(\theta_{\vb*{l}}) = \frac{\hbar\omega}{4\pi}g_{r}^{\mathrm{eff}} \theta_{\vb*{l}}^2.    
    \end{equation}
    Here we take into account that  $|\vb*{m}| \ll |\vb*{l}|$  in MnF$_2$, because the ratio of anisotropy and exchange magnetic fields is  $0.018$~\cite{Vaidya2020} which guarantees small $|\vb*{m}|$, so that contribution of $\vb*{m} \times \dot{\vb*{m}}$ term in Eq.~\eqref{eq:Is} is negligible. This might change in other AFIs, such as FeF$_2$, rendering Eq.~\eqref{eq:effectivesmc} inapplicable. The SMCs extracted from Eq.~\eqref{eq:effectivesmc}---which also carry the label LH or RH,  $g_{r}^{\mathrm{eff, LH}}$ and $g_{r}^{\mathrm{eff, RH}}$, depending on which of the two precession modes at AF resonance are excited~\cite{Vaidya2020}---are given in Table~\ref{tab:smc}. The values we compute for MnF$_2$/Pt clean interface are about an order of magnitude larger than \mbox{$\approx 2.86 \times 10^{18}$ m$^{-2}$} measured experimentally~\cite{Vaidya2020}.
    
    \begin{table}[t!]
    	\centering
    	\begin{tabular}{c|c|c }
    		\hline\hline
    		System & $g_{r}^{\rm eff,LH}$ & $g_{r}^{\rm eff,RH}$ \\ 
    		\hline
    		Cu/MnF$_2$/Cu & 15.82 & 27.89\\ \hline
    		Pt/MnF$_2$/Cu & 31.42 & 54.52\\ 
    		\hline\hline
    	\end{tabular}
    	\caption{The {\em effective} SMC $g_{r}^{\rm eff,LH}$ and $g_{r}^{\rm eff,RH}$ (in the units of $10^{18}~\mathrm{m}^{-2}$) for Cu/MnF$_2$ and Pt/MnF$_2$ interfaces is extracted from  Floquet-NEGF [Sec.~\ref{sec:floquetnegf}] combined with ncDFT calculations [Sec.~\ref{sec:floquethamiltonian}] of pumped spin current $I_{\mathrm{DC},L}^{S_z}$, which is then plugged into Eq.~\eqref{eq:effectivesmc} applicable for $\theta_{\vb*{l}} \le 10^\circ$ [Figs.~\ref{fig:fig3}(c) and ~\ref{fig:fig3}(d)].}
    	\label{tab:smc}
    \end{table}   
    
    \subsection{Comparison with spin pumping from 1D tight-binding model of AFI/NM interface and time-dependent NEGF formalism}\label{sec:tdnegf}
  
	Finally, we independently corroborate {\em unequal} [Fig.~\ref{fig:fig3}] spin currents pumped into the $L$ lead by LH vs. RH precession modes   
	of sublattice magnetizations $\vb*{m}_1(t)$ and $\vb*{m}_2(t)$ by applying fully time-dependent NEGF calculations~\cite{Gaury2014,Popescu2016,Petrovic2018} to a toy 1D model in Fig.~\ref{fig:fig4} where an infinite TB chain, attached to the $L$ and the $R$ macroscopic reservoirs kept at the same electrochemical potential, host classical precessing LMMs comprising AFI. The conduction electrons within TB chain interact with the first and the second site at each edge of AFI via the $sd$ exchange interaction, as described by the following spin and LMM-dependent, TB Hamiltonian~\cite{Petrovic2018,Suresh2020}
	\begin{equation}\label{eq:tbh}
		\hat{H}(t) = - \sum_{\langle ij \rangle} \gamma_{ij} \hat{c}_{i}^\dagger\hat{c}_j - J_\mathrm{sd} \sum_{i}\hat{c}_i^\dagger\boldsymbol{\sigma} \cdot \bold{M}_i(t) \hat{c}_i.
	\end{equation}
    Here for, e.g., RH mode we use $\mathbf{M}_i(t)=\big(\sin \theta \cos(\omega_0 t),\sin \theta \sin(\omega_0 t), \cos \theta \big)$
    and $\mathbf{M}_{i+1}(t)=\big(\sin \theta \cos(\omega_0 t + \pi),\sin \theta \sin(\omega_0 t + \pi), \cos \theta \big)$. Due to insulating nature of AFI, the nearest-neighbor (as signified by $\langle \ldots \rangle$) hoppings $\gamma_{ij}$ within it are: zero [denoted by no line between  sites in Fig.~\ref{fig:fig4}(a)]; \mbox{$\gamma_{ij}=\gamma=1$ eV} outside [denoted by solid line between  sites in Fig.~\ref{fig:fig4}(a)]; and we also include 
    smaller hopping $\gamma_{ij}=0.1 \gamma$  between the first and the second site [as denoted by dotted line between them in Fig.~\ref{fig:fig4}(a)] on each edge of AFI to include proximity effect (i.e., evanescent wave functions, see Fig.~\ref{fig:fig2}) of metallic $L$ and $R$ TB chains onto AFI in the middle. No SO coupling is included into $\hat{H}(t)$ in Eq.~\eqref{eq:tbh}.

	The quantum evolution of electrons is described by solving a matrix integro-differential equation for time dependence of the nonequilibrium density matrix~\cite{Popescu2016,Petrovic2018}
	\begin{equation}\label{eq:dm}
		i\hbar \frac{d {\bm \rho}_\mathrm{neq}}{dt} = [\mathbf{H}, {\bm \rho}_\mathrm{neq}] + i \sum_{p=L,R} [{\bm \Pi}_p(t) + {\bm \Pi}_p^\dagger(t)].
	\end{equation}
	This can be viewed as the exact master equation for an open finite-size quantum system, described by $\hat{H}(t)$ in Eq.~\eqref{eq:tbh} and its matrix representation $\mathbf{H}(t)$, that is attached (via semi-infinite NM leads) to macroscopic reservoirs. The matrices ${\bm \rho}_\mathrm{neq}(t)=\hbar \mathbf{G}^<(t,t')/i$ and ${\bm \Pi}_p(t)$ are expressed in terms of time-dependent NEGFs~\cite{Gaury2014,Stefanucci2013} or integrals [in the case of  ${\bm \Pi}_p(t)$] over them, as elaborated in Refs.~\cite{Popescu2016}. The ${\bm \Pi}_p(t)$ matrices yield generally time-dependent~\cite{Varella2021} charge
	\begin{equation}
	I_p(t)=\frac{e}{\hbar} \mathrm{Tr} \, [ {\bm \Pi}_p(t) ], 
	\end{equation}
	and spin
	\begin{equation}
	I_p^{S_\alpha}(t)=\frac{e}{\hbar} \mathrm{Tr}\, [ \hat{\sigma}_\alpha {\bm \Pi}_p(t)], 
	\end{equation}	
	currents pumped into the NM lead $p=L,R$.  

	The calculation of $I_p^{S_z}$, which is steady-state in the absence of SO coupling~\cite{Varella2021}, for 1D toy model of the AFI/NM interface in Fig.~\ref{fig:fig4} confirm  the difference between its LH and RH precession modes pumping of spin currents into the NM leads. In fact, these results  [Figs.~\ref{fig:fig4}(b) and ~\ref{fig:fig4}(c)] that are quite similar (independently of whether $L$ and $R$ leads are identical or not) to those [Fig.~\ref{fig:fig3}] obtained  for realistic Cu/MnF$_2$/Cu and Pt/MnF$_2$/Cu heterostructures described by much more complex ncDFT Hamiltonians.	
	
	%===================================================================================
	\section{Conclusions and Outlook}\label{sec:conclusions}
	%===================================================================================
	In conclusion, by combining the Floquet theory of periodically driven quantum systems~\cite{Shirley1965,Sambe1973} with time-dependent 
	NEGFs~\cite{Mahfouzi2012,Mahfouzi2014,Stefanucci2013,Gaury2014,Popescu2016} and ncDFT calculations~\cite{Capelle2001,Eich2013a}, we demonstrated  first-principles quantum transport methodology which makes it possible to compute {\em directly} time-averaged (i.e., the DC component of) pumped spin current by precessing magnetization in realistic AFI/HM heterostructures. The atomistic structure of such heterostructures, including strong interfacial SO coupling 
	and SO-proximitized AFI layer by the evanescent metallic states from HM layer, is accurately described by first-principles Hamiltonian as the only input in our microscopic calculations. Enhancement of SO coupling in AFI/HM heterostructure, by proper choice of HM layer, can {\em increase} pumped spin current and the corresponding effective SMC, which shows a pathway for materials design in order to tune the SMC for AF spintronic applications. In contrast, interfacial SO coupling  reduces~\cite{Dolui2020} the effective SMC in traditional~\cite{Tserkovnyak2005} FM/HM heterostructures or in recently explored FM/topological-insulator~\cite{Dolui2020,Jamali2015} heterostructures, although SMC remains of the same order of magnitude in all three types of magnetic heterostructures. Angular $\theta$ dependence pumped spin current $I^{S_z}$ by radiation-driven AFI is modulated by the handedness of the two AF modes, and consequently the corresponding effective SMC differs. Nevertheless, extraction of the effective~\cite{Dolui2020,Zhu2019} SMC is possible when $I^{S_z}$ exhibits $\propto \theta_{\vb*{l}}^2$ dependence for small $\theta_{\vb*{l}}$, as is the case of materials with $H_{\rm E} \gg H_{\rm A}$ exemplified by MnF$_2$. 

    We note that, in general, the Floquet theorem does not allow one to transform time-dependent DFT-Hamiltonian to a time-independent Floquet-DFT-Hamiltonian~\cite{Kapoor2013}. Instead, one apparently has to perform self-consistent time-dependent DFT calculations~\cite{Huebener2017} in the presence of time-periodic potential, in contrast to our first-principles Hamiltonian  where harmonic potential is introduced {\em a posteriori} into the converged KS Hamiltonian produced by {\em static} ncDFT calculations. Nevertheless, due to small  frequency, $\hbar \omega \ll E_F$, of electromagnetic radiation employed---in THz range for AFI~\cite{Vaidya2020,Li2020} or in GHz range for conventional FM---when compared to the Fermi energy of heterostructures in Fig.~\ref{fig:fig1}, we expect tiny perturbation of electronic density already converged by static ncDFT calculations. In other words, our much less expensive calculations are in the spirit of linear-response theory and first-principles scattering matrix calculation~\cite{Xia2002,Zwierzycki2005,Carva2007,Liu2014a} of SMC where one employs converged static KS Hamiltonian (but without SO coupling) as an input. 
    
    In the same limit $\hbar \omega \ll E_F$, our methodology can be applied to a plethora of problems outside of spintronics which involve Floquet engineering of quantum materials~\cite{Oka2019}. Also, it can accommodate impurities or thermal disorder---in the form of ``frozen-phonons'' and/or ``frozen-magnons''~\cite{Starikov2018}---directly through DFT Hamiltonian. This makes it possible to study temperature and disorder effects on SMC, at  additional computational cost when using supercells of larger size in the transverse direction and averaging over disorder configurations.

	\bigskip
	
	%======================================ACKNOWLEDGEMENTS=============================
	\begin{acknowledgments}
    K.~D. was supported by the U.S. Department of Energy (DOE) Grant No. DE-SC0016380. A.~S. and B.~K.~N. were supported by the US National Science Foundation (NSF) through the University of Delaware Materials Research Science and Engineering Center DMR-2011824. 
	\end{acknowledgments}
	%=======================================BIBLIOGRAPHY===============================
	%\bibliographystyle{unsrt}
	%\bibliographystyle{abbrvnat}
	%\bibliography{bibliography.bib}

		\end{document}